\documentclass[12pt]{article}
\usepackage{amsmath,latexsym}
\usepackage{graphicx}
\begin{document}

 \title{\Huge Wormhole and C-field:  Revisited}
 \author{F.Rahaman$^*$, M.Sarker$^*$ and M.Kalam$^{\ddag}$ }

\date{}
 \maketitle
 \begin{abstract}
Recently,  Rahaman et al [ Nuovo.Cim 119B, 1115(2004)]  have shown
that the static spherically symmetric solutions in presence of
C-field give rise to  wormhole geometry. We highlight some of
 the characteristics of  this wormhole, which have not been  considered
   in the  previous  study.
\end{abstract}

 %\bigskip
 %\medskip
  \footnotetext{ Pacs Nos :  04.20 Gz,04.50 + h, 04.20 Jb   
\\
 $*$Dept.of Mathematics, Jadavpur University, Kolkata-700 032, India

                                  E-Mail:farook\_rahaman@yahoo.com\\
$\ddag$Dept. of Phys. , Netaji Nagar College for Women, Regent Estate,
Kolkata-700092, India.\\

}
 %   \mbox{} \hspace{.2in}
%%Section Introduction
In recent times, the wormhole has become very popular as because
it could allow for interstellar distances to be traveled in very
short times. In a seminal paper, Morris and Thorne [MT][1] have
shown that wormholes are solutions of the Einstein's equations
that have two regions connected by a throat. They observed that it
is the solutions of Einstein's equations that shared  the
violation of null energy condition. This bizarre form of matter
that characterized the above stress energy tensor is known as
exotic matter. If, some how, an advanced engineers able to
manufacture the exotic matter, then it would be possible to
construct a wormhole. If wormhole could be constructed, the faster
than light travel would be possible. In other words, time machine
must be constructed. There are different ways of evading these
unexpected matter energy source. Most of these attempts focus on
alternative theories of gravity or phantom energy ( i.e. cosmic
fluid which is responsible for the accelerated expansion of the
Universe ). Several authors have explained wormholes in scalar
tensor theory of gravity in which scalar field may play the role
of exotic matter [2-7]. Last few years, physicists have discussed
the physical properties and characteristics of traversable
wormholes by taking phantom energy as source[8-12]. Long ago,
since 1966, Hoyle and Narlikar [HN] proposed an alternative theory
of gravity known as C-field theory [13]. HN adopted a field
theoretic approach introducing a massless and chargeless scalar
field C in the Einstein-Hilbert action to account for the matter
creation.

\pagebreak

The complete action functional describing C-field, matter and
gravity is taken as

 \begin{equation}A = \frac{1}{16 \pi G} \int R\sqrt{-g}d^4x  - \frac{1}{2} f
 \int C_iC^i \sqrt{-g}d^4x  + A_1 +  A_{matter}\end{equation}

Here $ C_i = ( \frac{\partial C}{\partial x_i}) $ and $f>0$ is a
coupling constant. The action $A_1$ can be taken to be $ \sum_n
\int C_i d^4x^i_n $ where the coordinates $x^i_n $ represent the
world line of the $n^{th}$ particle. On varying the above action
w.r.t. $g^{ab}$, one can get the usual Einstein equation with the
addition of C-field energy density
 \begin{equation}
              R^{ab} - \frac{1}{2}g^{ab}R
             = - 8\pi G [T^{ab}- f{ C^aC^b +
             \frac{1}{2}fg^{ab}C^iC_i}]
           \end{equation}
Here, $T_{ab}$ is the matter tensor.

 A C-field generated  by a
certain source equation, leads to interesting change in the
cosmological solution of Einstein field equations.
 Several authors,  have studied cosmological
models [14] and topological defects[15] in presence of C-field.
Since, C-field theory is a scalar tensor theory, so it is
interesting to search whether the creation field C may play the
role of exotic matter that is required to get wormhole solution.
Recently, Rahaman et al [16] have pointed out that a spherically
symmetric vacuum solutions to the C-field theory give rise to a
wormhole. In this report, we would like to mention some of the
characteristics ( i.e. matching with Schwarzschild metric,
traversability etc ) of the C-field wormhole, which have not been
considered in the previous  study.

Let us consider the static spherically symmetric metric as
\begin{equation}
               ds^2=  - e^\nu dt^2+ e^\mu dr^2+r^2( d\theta^2+sin^2\theta
               d\phi^2)
         \label{Eq3}
          \end{equation}

The independent field equations for the metric (3) are
\begin{equation}e^{-\mu}
[\frac{1}{r^2}-\frac{\mu^\prime}{r}]-\frac{1}{r^2}= 4\pi
Gfe^{-\mu}{(C^\prime)^2}\end{equation}
\begin{equation}e^{-\mu}
[\frac{1}{r^2}+\frac{\nu^\prime}{r}]-\frac{1}{r^2}= -4\pi
Gfe^{-\mu}{(C^\prime)^2}\end{equation}
\begin{equation}e^{-\mu}
[\frac{1}{2}(\nu^\prime)^2+ \nu^{\prime\prime}
-\frac{1}{2}\mu^\prime\nu^\prime + \frac{1}{r}({\nu^\prime-
\mu^\prime})] = 8\pi Gfe^{-\mu}{(C^\prime)^2}\end{equation}

\pagebreak

The solutions are given by [16]

\begin{equation}e^{\nu}= constant\end{equation}
\begin{equation}e^{-\mu}= 1-\frac{D}{r^2}\end{equation}
\begin{equation}C=\frac{1}{\sqrt(4\pi
Gf)}\sec^{-1}\frac{r}{\sqrt(D)}+ C_0\end{equation}

where D and $C_0$ are integration constants. We note that the
creation field becomes constant and equal to  $C_0$ when $ r
\rightarrow \infty $.

 Thus the metric (3) can be written in Morris-Thorne
cannonical form as

\begin{equation}
               ds^2=  - e^{\nu}dt^2+ \frac{dr^2}{[ 1-\frac{b(r)}{r}]}
               +r^2( d\theta^2+sin^2\theta
               d\phi^2)
          \end{equation}
Here,  $ b(r) =\frac{D}{r}$  is called shape function and $e^{\nu}
= $ redshift  function $   =  constant $ .

The  shape   function is depicted  in  fig.1.
\begin{figure}[htbp]
    \centering
        \includegraphics[scale=.8]{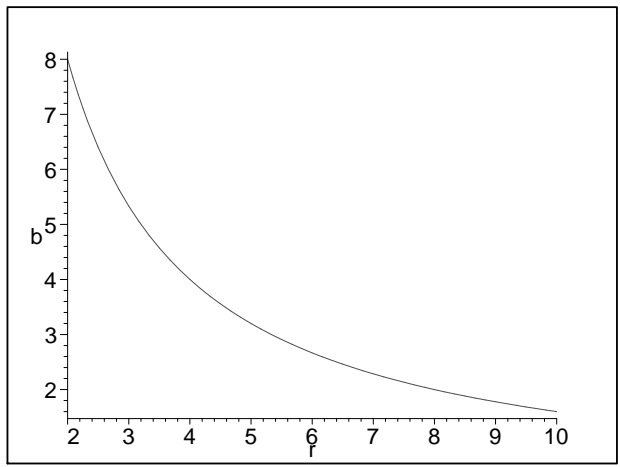}
        \caption{The shape function of the wormhole}
    \label{fig:cosh}
\end{figure}

We notice that the square root of the integration constant D
indicates the position where the throat of the
  wormhole occurs i.e. at $ r = r_0 =  \sqrt{D}>0 $. One
can note  that since now $r\geq r_0 > 0$, there is no horizon. Now
we match the interior wormhole solution to the exterior
Schwarzschild solution ( in the absence of C-field ). To match the
interior to the exterior, we impose the continuity of the metric
coefficients, $ g_{\mu\nu} $, across a surface, S , i.e. $
{g_{\mu\nu}}_{(int)}|_S =  {g_{\mu\nu}}_{(ext)}|_S $.

[ This condition is not sufficient to different space times.
However, for space times with a good deal of symmetry ( here,
spherical symmetry ), one can use directly the field equations to
match. Actually,  if the metric  coefficients are not
differentiable and affine connections are not continuous at the
junction then one has to use the second fundamental forms
associated with the two sides of the junction surface[17] ]

The wormhole metric is continuous from the throat, $ r = r_0$ to a
finite distance $ r = a $. Now we impose the continuity of $
g_{tt} $ and $ g_{rr}$,

$ {g_{tt}}_{(int)}|_S =  {g_{tt}}_{(ext)}|_S $

$ {g_{rr}}_{(int)}|_S =  {g_{rr}}_{(ext)}|_S $

at $ r= a $ [ i.e.  on the surface S ] since $ g_{\theta\theta} $
and $ g_{\phi\phi}$ are already continuous.

The continuity of the metric then gives generally

$ {e^{\nu}}_{int}(a) = {e^{\nu}}_{ext}(a) $ and $
{e^{\mu}}_{int}(a) = {e^{\mu}}_{ext}(a) $.

Hence one can find
\begin{equation}e^{\nu}= ( 1 - \frac{2GM}{a}) \end{equation}
and $  1 - \frac{b(a)}{a} = ( 1 - \frac{2GM}{a})  $ i.e. $ b(a) =
2GM $

This implies $ \frac{D}{a} = 2GM $

Hence,
\begin{equation} a = \frac{D}{2GM} \end{equation}
i.e. matching occurs at $  a = \frac{D}{2GM} $.

The interior metric $ r_0 < r \leq a $ is given by

\begin{equation}
               ds^2=  - [ 1-\frac{D}{a^2}]dt^2+ \frac{dr^2}{[ 1-\frac{D}{r^2}]}
               +r^2( d\theta^2+sin^2\theta
               d\phi^2)
          \end{equation}

The exterior metric $ a \leq r < \infty   $ is given by

\begin{equation}
               ds^2=  - [ 1-\frac{D}{ar}]dt^2+ \frac{dr^2}{[ 1-\frac{D}{ar}]}
               +r^2( d\theta^2+sin^2\theta
               d\phi^2)
          \end{equation}

Since wormhole is not a black hole, so we have to impose the
condition $ a > 2GM $.

In recent, Das and Kar [18] have done an interesting work where
they have shown that this geometry ( corresponding to eq.13 ) ,
can also be obtained with tachyon matter as a source term in the
field equations and a positive cosmological constant.

The axially symmetric embedded surface $ z = z(r)$ shaping the
Wormhole's spatial geometry is a solution of

\begin{equation}\label{Eq21}
 \frac{dz}{dr}=\pm \frac{1}{\sqrt{\displaystyle{\frac{r}{b(r)}}-1}}.
 \end{equation}

  One can note from the definition of wormhole that at   $ r= r_0 $ (the wormhole throat)
  eq.15 is divergent i.e.  embedded surface is vertical there.

The embedded surface (solution of eq.15) in this case is ,

\begin{equation}\label{Eq22}
z =  \sqrt{D} \cosh^{-1}   \frac{r}{\sqrt{D}}.
\end{equation}

\begin{figure}[htbp]
    \centering
        \includegraphics[scale=.8]{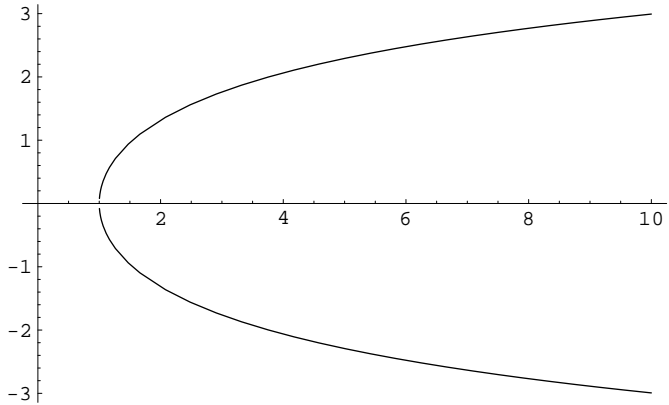}
        \caption{The embedding diagram of the wormhole}
    \label{fig:cosh}
\end{figure}

\pagebreak

One can see embedding diagram of this wormhole in Fig.2. The
surface of revolution of this curve about the vertical z axis
makes the diagram complete ( see Fig.3 ).

\begin{figure}[htbp]
    \centering
        \includegraphics[scale=.8]{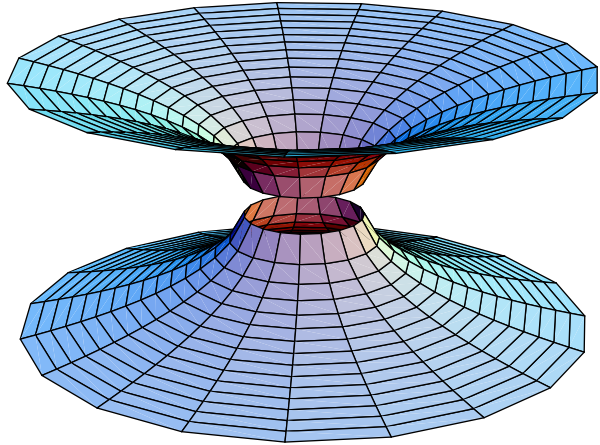}
    \caption{The full visualization of the surface generated by the rotation of the
     embedded
    curve about the vertical z axis}
    \label{fig:wormhole}
\end{figure}

According to Morris and Thorne [1], the '$r$' co-ordinate is
ill-behaved near the throat, but proper radial distance
\begin{equation}\label{Eq23}
 l(r) = \pm \int_{r_0^+}^r \frac{dr}{\sqrt{1-\frac{b(r)}{r}}}
 \end{equation}
 must be well behaved everywhere i.e. we must require that $ l(r) $ is finite
 throughout the space-time.

 In this model,

\begin{equation}\label{Eq24}
l(r) =  \pm\sqrt {r^2 - D }.
 \end{equation}

This is a well behaved coordinate system. The radial distance is
positive above the throat (our Universe) and negative below the
throat (other Universe). At very large distance from the throat,
the embedding surface becomes flat $\displaystyle{\frac{dz}{dr}}(
l \longrightarrow  \pm  \infty ) = 0 $ corresponding to  the two
asymptotically flat regions ($l \longrightarrow+\infty $ and $l
\longrightarrow- \infty$), which the wormhole connects.

\begin{figure}[htbp]
    \centering
        \includegraphics[scale=.8]{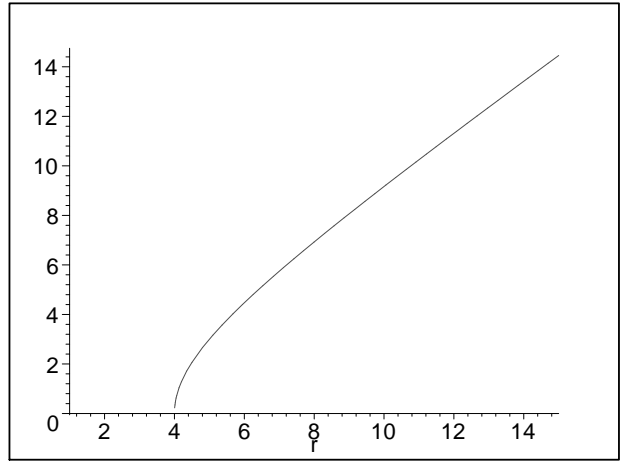}
        \caption{Diagram of the radial proper distance ( D = 16)( upper half)}
   \label{fig:shape4}
\end{figure}

\pagebreak

The radial proper distance is measured from $r_0$ to any $ r >
r_0$. Note that on the throat $r = r_0$,  $ l= 0$.

Now we will focus to the traversability condition of a human
being. It is necessary that the tidal accelerations between two
parts of the traveller's body, separated by say, 2 meters, must
less than the gravitational acceleration at earth's surface
$g_{earth}$ (  $ g_{earth} \approx 10 m / s^2 $ ).

 According  to MT [1], one obtains the following
inequality for tangential tidal acceleration ( assuming $
\nu^\prime = 0 $ ),

\begin{equation}\label{Eq24}
 \mid \frac{\beta^2}{2r^2} (\frac{v}{c} )^2( b^\prime - \frac{b}{r})|
 \leq  \frac{g_{earth}}{2c^2 m } \approx \frac{1}{10^{16} {m^2}}
 \end{equation}

with $ \beta = \frac{1}{ \sqrt{1 - (\frac{v}{c} )^2}}$ and v being
the traveller's velocity.

For  $ v << c $, we have $\beta \approx 1 $ and substituting the
expression b(r) given in (10), one gets,

\begin{equation}\label{Eq24}
 \frac{v}{c} < \frac{r^2}{\sqrt{D}10^8 m}
 \end{equation}

If we consider the traveller velocity $ v = .005c $ at the throat
$ r = r_0 $, one finds, $ r_0  \approx 5 \times 10^5 m $.

Taking into account equation (12) for asymptotically flat
spacetimes, the region of matter distribution will extend to $ a =
\frac{25 \times 10^{10}}{ 2GM} $.

One can note that, by choosing a, we find the value of the
wormhole mass.

The radial tidal acceleration is zero since $ \nu^ \prime = 0 $.

Acceleration felt by a traveller should less than the
gravitational acceleration at earth surface,  $g_{earth}$. The
condition imposed by MT as [ for $ \nu^\prime = 0 $]

\begin{equation}\label{Eq24}
 |\textbf{f}| = |\sqrt{[ 1 - \frac{b(r)}{r}}] \beta ^ \prime c^2| \leq  g_{earth}
 \end{equation}

 For the traveller's velocity $ v = constant$, one finds that $
 |\textbf{f}|= 0
 $.
 In our model the condition (21) is automatically satisfied, the
 traveller feels a zero gravitational acceleration.

Now we consider, the trip takes the time say less or equal to one
year for both the traveller and the observer that stay at rest at
the space stations $ l = - l_1 $ and $ l= + l_2 $ as [1]

$ \Delta\tau_{ traveler} = \int_{-l_1}^{l_2} \frac{dl}{v\beta}
\leq 1 $ year$  $

$ \Delta t_{ space station} = \int_{-l_1}^{l_2}
\frac{dl}{\sqrt{g_{tt}}v} \leq 1 $ year$  $.

 For low velocity $ v << c $, we have $ \beta
\approx 1 $ and with $ e^{\nu}= constant $, the above expressions
reduce to

$ \Delta\tau_{ traveler} \approx \frac{2l}{v\beta} =
\sqrt{g_{tt}}\Delta t_{ space station} $.

For $ a >> r_0 $, $ g_{tt} = ( 1 - \frac{D}{a^2}) \approx 1 $, so
that $ \Delta\tau_{ traveler} \approx \Delta t_{ space station}
\approx  \frac{2a}{v} $.

Here $ \frac{2a}{v} \approx  1 $ year $ = 3.16  \times 10^7 s $.

If the traversal velocity is $ v = .005c $, the junction surface
is at $ a = 7.9 \times 10^{10}  m $.

In conclusion, we have discussed some characteristics of the
spherically symmetric wormhole solution in presence of C-field. We
have established a matching of an interior solution with an
exterior Schwarzschild solutions. How a massive body warps space
would be visualized by the embedding a curved two dimensional
surface in a three dimensional flat ( Euclidean ) space. In our
model, the wormhole can be visualized as catenoid of revolution $
r = \sqrt{D}$ cosh$ \frac{z}{\sqrt{D}}$. According to MT [1], if
wormhole travel is possible for human beings, the traveller's
journey must satisfy three constraints: (i) the entire trip should
require less than or of order 1 year as measured both by the
traveller and by the peoples who live in the stations at $ l = l_2
$ and $ l = - l_1 $ (ii) the acceleration \textbf{f}, felt by the
traveller less than 1 earth gravity (iii) the tidal accelerations
between various parts of the traveller's body less than 1 earth
gravity. In our model all the three conditions have been verified.
Thus the C-field generated wormholes are usable and traversable in
practice.

\pagebreak

One should be noted that there exists some regions in which
C-field may play the role of exotic matter. The creation field C
is depicted in fig.5.

\begin{figure}[htbp]
    \centering
        \includegraphics[scale=.8]{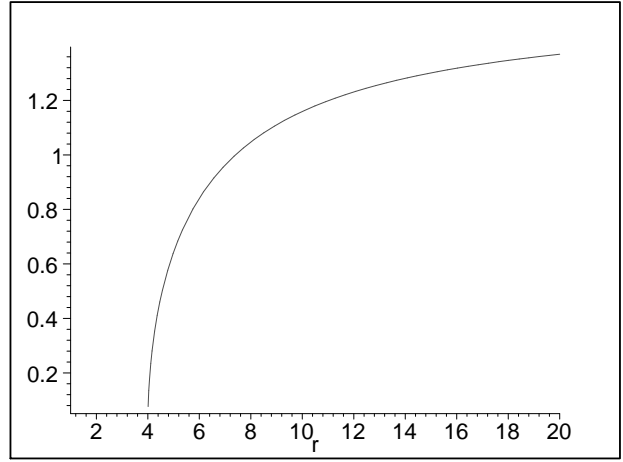}
        \caption{The diagram for the creation field C }
   \label{fig:shape4}
\end{figure}

Recently, a measure of quantifying exotic matter needed for
traversable wormhole geometry has been developed within the
framework of general relativity [19-20]. It is interesting to
mention that no such quantifying measure is really needed in our
model.

{ \bf Acknowledgments }

          F.R is thankful to DST , Government of India for providing
          financial support. MK has been partially supported by
          UGC,
          Government of India under MRP scheme. We are grateful to the referee
          for his valuable comments.\\


\begin{thebibliography}{99}
\bibitem{kg6}  M. Morris and K. Thorne , American  J. Phys. 56, 39 (1988 )

\bibitem{kg1} A Bhadra and K Sarkar, arXiv:gr-qc/0503004
    \bibitem{kg1} K K Nandi et al, Phys. Rev. D 57, 823 (1997)
    \bibitem{kg1}L
    Anchordoqui et al, Phys. Rev. D 55, 5226 (1997)
    \bibitem{kg1} A Agnese and M
    Camera, Phys. Rev. D 51, 2011(1995)
    \bibitem{kg1} K K Nandi et al, Phys. Rev. D 55, 2497 (1997)

\bibitem{kg1}F Rahaman, M Kalam and  A Ghosh, arXiv:gr-qc/0605095
\bibitem{kg10} F. Lobo,  arXiv: gr-qc/0502099; arXiv: gr-qc/0506001
; arXiv: gr-qc/0603091
\bibitem{kg10} S. Sushkov,  arXiv: gr-qc/0502084

 \bibitem{kg10}O. Zaslavskii,  arXiv: gr-qc/0508057

  \bibitem{kg10}F Rahaman, M Kalam, M Sarker and K Gayen,
  gr-qc/0512075;
  F Rahaman et al, gr-qc/0611133;  F Rahaman et al, gr-qc/0701032
  ;  F Rahaman et al, Phys.Scr. 76, 56 (2007) (arXiv:0705.1058 [gr-qc]  )
  \bibitem{kg10} V Faroni and W Israel, arXiv: gr-qc/0503005
  \bibitem{kg4}  Hoyle. F and Narlikar. J. V., Proc.Roy.Soc. A 290 (1966) 162


\bibitem{kg7}  F Rahaman, B C Bhui and P Ghosh, Nuovo
Cim.119B:1115-1119,2004; e-Print: gr-qc/0512113
\bibitem{kg8}  Narlikar. J and Padmanabhan.T, Phys.Rev.D 32, 1928 (1985);
S.Chaterjee and A  Banerjee, Gen.Rel.Grav.36:303-313,2004
\bibitem{kg8}F Rahaman  et al , Astrophys.Space Sci.302, 171(2006);
F Rahaman  et al, Chin.J.Phys.43, 806(2005); F Rahaman et al,
Int.J.Mod.Phys.A21:3727-3732,2006; e-Print: gr-qc/0601005
\bibitem{kg8}A Taub, J Math. Phys. 21, 1423 (1980); J Lemos, F
Lobo and S Oliveira, arXiv: gr-qc/0302049; F Rahaman et al,
arXiv:0705.0740 [gr-qc]
\bibitem{kg8} A Das and S Kar, arXiv: gr-qc/0505124
\bibitem{kg8}  M Visser, S Kar and N Dadhich,  arXiv: gr-qc/0301003
\bibitem{kg10}P Kuhfittig, arXiv: gr-qc/0401048


\end{thebibliography}
\end{document}